\documentclass[11pt]{article}\topmargin=-0.7cm \textheight 222mm \textwidth 15.8cm\oddsidemargin=2mm 
\newcommand{\comm}[1]{}

\usepackage{amssymb}
\usepackage{amsmath}\usepackage{graphicx} \usepackage[numbers]{natbib}\usepackage{epsfig}
\def\citet{\cite}

\def\xxxonly{\comm}

\usepackage{times}\hfuzz=10pt 

 \usepackage{color}

\newtheorem{theorem}{Theorem}
\newtheorem{lemma}{Lemma}
\newtheorem{corollary}{Corollary}
\newtheorem{definition}{Definition}
\newtheorem{remark}{Remark}
\newtheorem{example}{Example}

\def\e{\varepsilon}

\def\defi{\stackrel{{\scriptscriptstyle \Delta}}{=}}

\def\OO{{\scriptscriptstyle O}}

\def\a{\alpha}
\def\d{\delta}
\def\o{\omega}
\def\O{\Omega}

\def\F{{\cal F}}
\def\w{\widehat}
\def\Ind{{\mathbb{I}}}

\def\Re{{\rm Re\,\!}}

\def\R{{\bf R}}

\def\Z{{\cal Z}}
\def\ZZ{{\bf Z}}

\def\s{\delta}
\def\g{\gamma}
\def\C{{\bf C}}

\def\ww{\widetilde}

\def\t{\theta}
\def\oo{\bar}
\def\s{\sigma}

\def\U{{\cal U}}
\def\V{{\cal V}}

\def\M{{\cal M}}

\def\T{{\mathbb{T}}}

\newcommand{\be}{\begin{equation}}
\newcommand{\ee}{\end{equation}}
\newcommand{\bd}{\begin{displaymath}}
\newcommand{\ed}{\end{displaymath}}
\newcommand{\ba}{\begin{array}{ll}}
\newcommand{\ea}{\end{array}}
\newcommand{\baa}{\begin{eqnarray}}
\newcommand{\eaa}{\end{eqnarray}}
\newcommand{\baaa}{\begin{eqnarray*}}
\newcommand{\eaaa}{\end{eqnarray*}}
\font\sm=cmr10

\def\oo{\bar}

\def\a{\alpha}

\def\ko{k}
\def\yo{y}
\def\ew{\left(e^{i\o}\right)}
\def\BL{{\scriptscriptstyle BL}}
\def\BLO{L_{2}^{\BL,\O}(\R)}
\def\BLOO{L_{2}^{\BL,\pi/\tau}(\R)}
\def\T{{\mathbb{T}}}
\def\ZZ{{\mathbb{Z}}}
\def\NN{{\mathbb{N}}}
\def\TT{{\cal T}}

\def\HHH{{\rm H}}

\def\ee{\epsilon}
\def\V{V}
\date{Submitted: May 2, 2016. Revised: October 29, 2017}
\title{
On sampling theorem with sparse decimated  samples: exploring branching spectrum degeneracy}
\author{
Nikolai Dokuchaev}
 
\begin{document}
 \vspace{-0.5cm}      \maketitle
\def\brea{}
\def\breakk{}
\def\break{}
\begin{abstract}
The paper investigates possibility of recovery of sequences from their sparse decimated subsequences.
It is shown that  this recoverability is associated with certain spectrum degeneracy of a new kind (branching degeneracy), and that sequences of a general kind can be approximated by sequences with this property.
  Application of this result  to equidistant samples of continuous time band-limited functions
   featuring this degeneracy allows to establish recoverability of these functions from sparse subsamples below the critical
     Nyquist rate.
\index{in
IEEE:}\let\thefootnote\relax\footnote{The author is with  Department of
Mathematics and Statistics, Curtin University, GPO Box U1987, Perth,
Western Australia, 6845 (email N.Dokuchaev@curtin.edu.au).  \index{This work  was
supported by ARC grant of Australia DP120100928 to the author.}}

\end{abstract}\section{Introduction}
The paper investigates possibility of recovery of sequences from their decimated subsequences.
It appears there this recoverability is associated with certain spectrum degeneracy of a new kind, and that a
sequences of a general kind can be approximated by sequences featuring this degeneracy.
This opens some opportunities for sparse sampling of continuous time band-limited functions.
 As is known, the sampling rate required to recovery of  function is defined by the classical Sampling Theorem
 also known as
Nyquist-Shannon theorem, Nyquist-Shannon-Kotelnikov theorem,   Whittaker-Shannon-Kotelnikov theorem, Whittaker-Nyquist-Kotelnikov-Shannon theorem,
which is one of the most basic results in the theory of signal processing and information
science;  the result  was obtained independently by four  authors  \citet{Whit,Nyq,Kotel,Shannon}. The theorem states that
 a band-limited function can be uniquely recovered without error  from  a infinite two-sided equidistant
sampling sequence   taken with sufficient frequency:  the sampling rate must be at least twice the maximum frequency
present in the signal (the critical Nyquist rate). Respectively,  more frequent sampling is required for a larger  spectrum domain. If the sampling rate is preselected, then it is impossible to approximate a function of a general type by a band-limited function that is uniquely defined by its sample with this particular  rate.
 This
principle defines the choice of the sampling rate in almost all signal processing protocols.
A similar principle works for discreet time processes and defines  recoverability of sequences from their subsequences. For example, sequences
with a spectrum located inside the interval  $(-\pi/2,\pi/2)$ can be recovered from their
decimated subsequences consisting of all terms with even numbers allows to recover sequences.

 Numerous  extensions of the sampling theorem  were obtained,
 including
the case of nonuniform sampling and restoration of the signal with mixed samples;
see  \citet{BM,Cai,jerri,F95,La,La2,OU08,U1,U2,U50,V87,V01,Z} an literature therein. There were works studying
 possibilities to reduce  the set of sampling points
required for restoration  of the underlying functions.
In particular, it was found   that a band-limited function can be recovered without error from an oversampling sample sequence  if a finite number of sample values is unknown, or if an one-sided half of any oversampling sample sequence is unknown \citet{V87}.
It was found   \citet{F95}
that the function can be recovered without error  with a missing equidistant  infinite subsequence consistent of  $n$th
member of the original sample sequence, i.e. that each $n$th member is redundant, under some additional constraints on the
oversampling parameter. The constraints are such that the oversampling parameter is increasing if  $n\ge 2$ is decreasing. There is also an approach based on the so-called Landau's phenomenon \cite{La,La2};
see \cite{BM,La,La2,OU08,U1,U2} and a recent literature review in \cite{OU}.
This approach
allows  arbitrarily sparse discrete uniqueness sets in the time domain for a fixed spectrum range; the focus is on the uniqueness problem rather than on algorithms for recovery.
Originally, it was shown in  \cite{La} that the set of sampling points representing small deviations of integers is  an uniqueness set for classes of functions with an arbitrarily large measure of the spectrum range, given that there are periodic gaps in the spectrum and that the spectrum range is bounded. The implied sampling points were not equidistant and have to be calculated.
This result was significantly extended. In particularly,    similar uniqueness for functions with unbounded spectrum range and for  sampling points
 allowing a simple explicit form  was established in \cite{OU08} (Theorem 3.1 therein). Some generalization  for multidimensional case were obtained in \cite{U1,U2}.
However, as was emphasized in \cite{La2},  the  uniqueness theorems  do not ensure  robust  data recovery; moreover, it is also known that any sampling below the Nyquist rate  cannot be stable in this sense \cite{La2,U2}.

 The present paper readdresses the problem of  sparse
 sampling using a different approach. Instead of analyzing continuous time functions with spectrum degeneracy,  it focuses  on analysis of spectrum characteristics of  sequences
that can be recovered from their periodic subsequences,  independently from the sampling problem for the continuous time functions.
  The goal was to describe for a given integer $m>0$, class of sequences $\ww x\in\ell_2$,
featuring the following properties:
 \begin{itemize}
\item[(i)]  $\ww x$ can be recovered from a subsample $\ww x(km)$;
\item[(ii)] these processes are everywhere dense in $\ell_2$, i.e. they can approximate  any  $x\in\ell_2$.
\end{itemize}
\par

We found
a solution based on a special kind of
spectrum degeneracy. This degeneracy does not assume that there are spectrum gaps for a frequency characteristic such as Z-transform.

Let us describe briefly introduced below classes of processes with these properties.  For a process
 $x\in \ell_2$, we consider an auxiliary  "branching" process, or a set of processes  $\{\w x_d\}_{d=-m+1}^{m-1}\in \ell_2$
 approximating alterations of the original process.
It appears that certain conditions of periodic degeneracy of the spectrums for all $\w x_d$,
  ensures that it is possible to compute a new representative sequence  $\ww x$ such that there exists a  procedure for recovery $\ww x$ from the subsample $\{\ww x(km)\}$. The procedure
  uses  a  linear predictor  representing a modification of the predictor from \cite{D12a} (see the proof of  Lemma \ref{ThOdd} below). Therefore,  desired properties (i)-(ii) hold.
  We interpret is as  a spectrum degeneracy of a new kind for $\ww x$  (Definition \ref{defAD1} and Theorems \ref{ThDeg}--\ref{ThD2} below).
In addition, we show  that the procedure of recovery of any finite part of the
  sample  $\ww x(k)$ from the subsample $\ww x(km)$ is robust with respect to noise contamination and data truncation, i.e. it
  is  a well-posed problem in a certain sense (Theorem \ref{ThA}).

Further, we apply the results sequences to  the
sampling of  continuous time band-limited functions.  We  consider  continuous time band-limited functions which samples are sequences
  $\ww x$ featuring the degeneracy  sequences mentioned above. As a corollary, we found that  these functions are uniquely defined by $m$-periodic subsamples
   of their equidistant  samples with  the critical Nyquist rate  (Corollary \ref{corr1}  and Corollary \ref{ThAC}).
 This allows to bypass, in a certain sense, the restriction on the sampling rate described by the critical   Nyquist rate, in the sprit of \cite{BM,La,La2,OU08,U1,U2}.
 The difference is that  we consider equidistant sparse sampling; the sampling points in \cite{BM,La,La2,OU08,U1,U2} are not equidistant, and this was essential. Further, our method is based on a recovery
 algorithm for sequences;  the results in \cite{BM,La,La2,OU08,U1,U2} are for the uniqueness problem and do not cover recovery algorithms.
 In addition, we consider different classes of functions; Corollary \ref{corr1}  and Corollary \ref{ThAC}
 cover band-limited functions, on the other hands, the uniqueness result
 \cite{OU08} covers function with unbounded spectrum periodic gaps. These gaps are not required
 in Corollary \ref{corr1}; instead,we request that sample series for the underlying function featuring spectrum degeneracy according to Definition \ref{defAD1} below.

\subsubsection*{Some definitions}
We denote by $L_2(D)$ the usual Hilbert space of complex valued
square integrable functions $x:D\to\C$, where $D$ is a domain.

We denote by $\ZZ$  the set of all integers. Let $\ZZ^+=\{k\in\ZZ:\ k\ge 0\}$ and let $\ZZ^-=\{k\in\ZZ:\ k\le 0\}$.
\par
For a set $G\subset \ZZ$ and $r\in[1,\infty]$,
we denote by $\ell_r(G)$ a Banach
space of complex valued sequences $\{x(t)\}_{t\in G}$
such that
$\|x\|_{\ell_r(G)}\defi \left(\sum_{t\in G}|x(t)|^r\right)^{1/r}<+\infty$ for $r\in[1,+\infty)$,
and $\|x\|_{r(G)}\defi \sup_{t\in G}|x(t)|<+\infty$ for $r=\infty$.
We denote  $\ell_r=\ell_r(\ZZ)$.

\par
Let $D^c\defi\{z\in\C: |z|> 1\}$, and let $\T=\{z\in\C: |z|=1\}$.
\par
For  $x\in \ell_2$, we denote by $X=\Z x$ the
Z-transform  \baaa X(z)=\sum_{k=-\infty}^{\infty}x(k)z^{-k},\quad
z\in\C. \eaaa Respectively, the inverse Z-transform  $x=\Z^{-1}X$ is
defined as \baaa x(k)=\frac{1}{2\pi}\int_{-\pi}^\pi
X\left(e^{i\o}\right) e^{i\o k}d\o, \quad k=0,\pm 1,\pm 2,....\eaaa
For $x\in \ell_2$, the trace $X|_\T$ is defined as an element of
$L_2(\T)$.

For $f\in  L_2(\R)$, we denote by $F=\F f$ the function
defined on $i\R$ as the Fourier transform of $f$;
$$F(i\o)=(\F f)(i\o)= \int_{-\infty}^{\infty}e^{-i\o t}f(t)dt,\quad
\o\in\R.$$ Here $i=\sqrt{-1}$ is the imaginary unit. For $f\in L_2(\R)$, the
Fourier transform $F$ is defined as an element of $L_2(i\R)$, i.e.
 $F(i\cdot)\in L_2(\R)$).

For $\O>0$, let $\BLO$ be the subset of $L_2(\R)$ consisting of
functions $f$  such that $f(t)=(\F^{-1}
F)(t)$, where $F(i\cdot)\in L_2(i\R)$ and  $F(i\o)=0\ \hbox{for}\ |\o|>\O$.

 Let  $\HHH^r(D^c)$ be the Hardy space of functions that are holomorphic on
$D^c$ including the point at infinity  with finite norm
$\|h\|_{\HHH^r(D^c)}=\sup_{\rho>1}\|h(\rho e^{i\o})\|_{L_r(-\pi,\pi)}$, $r\in[1,+\infty]$.
\index{Note that Z-transform defines a bijection
between the sequences from $\ell_2^+$ and the restrictions (i.e.,
traces) $X|_{\T}$ of the functions from $\HHH^2(D^c)$ such that  $\overline{X\ew} =X\left(e^{-i\o}\right)$ for $\o\in\R$; see, e.g., \cite{Linq}, Section 4.3.
If $X\ew\in L_1(-\pi,\pi)$ and $\overline{X\ew} =X\left(e^{-i\o}\right)$, then $x=\Z^{-1}X$
is defined as an element of $\ell_\infty^+$.}
\section{The main results}\label{SecM}
Up to the end of this paper, we assume that we are given $m\ge 1$, $m\in\ZZ$.

\subsection*{A special type of spectrum degeneracy for sequences}
We consider first some problems of recovering sequences from their decimated subsequences.
\begin{definition}\label{defBranch}  Consider an ordered set $\{ x_d\}_{d=-m+1}^{m-1}\in (\ell_2)^{2m-1}$
such that $ x_d(k)= x_0(k)$ for $k\le 0$ and $d>0$, and that $ x_d(k)= x_0(k)$ for $k\ge 0$ and $d<0$. We say that this is a branching process, and we say that
  $ x_0$ is its root.
\end{definition}
\begin{definition}\label{defMC} Consider a branching process
$\{x_d\}_{d=-m+1}^{m-1}\in (\ell_2)^{2m-1} $. Let  $\ww x\in \ell_2$ be defined such that  $\ww x(k)= x_d(k+d)$,
where $d$ is such that $ (k+d)/m\in\ZZ$ and that $d\in \{0,1,...,m-1\}$ if $k\ge 0$,
 $d\in \{-m+1,...,0\}$ if $k< 0$.
Then $\ww x$ is called the representative  branch for this branching process.
\end{definition}

For $\d>0$ and  $n\ge 1$, $n\in\ZZ$,  let \baa
 &&s_{n,k}=\frac{2\pi k-\pi}{ n}, \qquad k=0,1,...,n-1,\quad\nonumber\\
  &&J_{\d,n}=\{\o\in (-\pi,\pi]: \min_{k=0,1,...,n-1}|\o-s_{m,k}|\le \d\}.\qquad
\label{J}\eaa

Let $\V(\d,n)$ be the set of all   $x\in L_2(\R)$ such that
$X\ew=0$ for $\o\in J_{\d,n}$, where $X=\Z x$.

\par
 For $d=-m+1,...,m-1$, let us select   positive integers $\zeta(d)$ and
 $\mu(m,d)=m\zeta(d)$ such that
 \index{$s_{\mu(m,d_1),k}\neq s_{\mu(m,d_2),l}$ if $d_1\neq d_2$ for all $d_1,d_2\in\{-m+1,....,m-1\}$
 and $k\in\{0,1,...,\mu(m,d_1)-1\}$, $l\in\{0,1,...,\mu(m,d_2)-1\}$. In this case, }
 the sets
 $J_{\d, \mu(d,m)}$ are mutually disjoint for $d=-m+1,....,m-1$ for sufficiently small $\d>0$.
 \begin{example}\label{ex1} A possible choice of $\mu(d,m)$ is  $\mu(m,d)\defi 2^d m$ for $d\ge 0$ and   $\mu(m,d)=2^{2m+d-1} m$ for $d<0$.
 \end{example}

\begin{definition}\label{defAD}  Let $\d>0$  be given. We say that a branching process
$\{\w x_d\}_{d=1-m}^{m-1}\in (\ell_2)^{2m-1}$ features spectrum $(\d,m)$-degeneracy  if
  $\w x_d\in \V(\d, \mu(d,m))$ for all $d$.   We denote by $ \U(\d,m)$ the set  of  all branching processes with this feature.
  \end{definition}
\begin{definition}\label{defAD1}  Let $\ww x\in\ell_2$ be a representative branch
of a branching process from $\U(\d,m)$ for some $\d>0$.  We say that $\ww x$
features branching spectrum $(\d,m)$-degeneracy with parameters $(\d,m)$.
 We denote by $U(\d,m)$ the set  of  all sequences $\ww x$ with this feature.
 \end{definition}
\begin{theorem}\label{ThDeg}
\begin{enumerate}
\item
For any $\d>0$, $\ww x\in \cup_{\d>0}U(\d,m)$, and  $s\le 0$, the sequence
$\{\ww x(k)\}_{k\in\ZZ,\, k>0}$ is uniquely defined by the sequence
$\{\ww x(mk)\}_{k\in \ZZ,\, k\le s}$.
\item
For any $\d>0$,  $\ww x\in \cup_{\d>0}U(\d,m)$, and $s\ge 0$, the sequence
$\{\ww x(k)\}_{k\in\ZZ,\, k<0}$ is uniquely defined by the sequence
$\{\ww x(mk)\}_{k\in \ZZ,\, k\ge s}$.
\item
For any $s>0$,  any $\ww x\in \cup_{\d>0}U(\d,m)$  is uniquely defined by the subsequence   $\{\ww x(mk)\}_{k\in \ZZ:\  |k|\ge s}$.
 \end{enumerate}
\end{theorem}


\begin{theorem}\label{ThDense}
For any branching process
$\{x_d\}_{d=-m+1}^{m-1}\in (\ell_2)^{2m-1}$, $m\ge 1$, $m\in\ZZ$,  and any $\e>0$, there exists  a branching process
$\{\w x_d\}_{d=-m+1}^{m-1}\in \cup_{\d>0} \U(\d,m)$  such that \baa
\max_d\|x_d-\w x_d\|_{\ell_2}\le \e.
\label{xdd}\eaa
\end{theorem}

\par
Consider mappings $\M_d:\ell_2\to \ell_2$
such that, for  $x\in\ell_2$,  the sequence $x_d=\M_d x$ is defined as the following:
 \baaa
&&\hphantom{}\hbox{(a)}\,\,\,\, x_0=x; \hphantom{xxxxx}  \hphantom{x_d(k)=x(0),\quad k= 0,1,...,d,\quad k<0}\,\,\nonumber \\
&&\hphantom{}\hbox{(b)\,\, for} \quad d>0:\quad \breakk \hphantom{xxxxx}  x_d(k)=x(k),\quad k<0, \nonumber\\
&&\hphantom{xxxxx} x_d(k)=x(0),\quad k= 0,1,...,d, \nonumber\\
&&\hphantom{xxxxx} x_d(k)=x(k-d),\quad k\ge d+1;\nonumber\\
&&\hphantom{}\hbox{(c) \,\, for} \quad d<0:\quad \breakk\hphantom{xxxxx}  x_d(k)=x(k),\quad k>0, \nonumber\\
&&\hphantom{xxxxx} x_d(k)=x(0),\quad k= 0,-1,...,d, \nonumber\\
&&\hphantom{xxxxx} x_d(k)=x(k+d),\quad k\le d-1,....\hspace{-1cm}\label{xd}
\eaaa
\par Let $x\in\ell_2$, and let a branching  process
$\{\w x_d\}_{d=-m+1}^{m-1}$ be such that  $x_d=\M_d x$. It follows from the definitions that, in this case, the root branch $x_0$ is also the representative branch.  In this case, Theorem \ref{ThDeg} implies that the sequence  $x_0$ is uniquely defined by its subsequence  $\{x_0(mk)\}_{k\in\ZZ}$.

\begin{theorem}\label{ThD2}
For any $x\in\ell_2$ and any $\ww\e>0$, there exists  $\ww x\in\cup_{\d>0}U(\d,m)$  such that \baa
\|x-\ww x\|_{\ell_2}\le \ww\e.
\label{xd1}\eaa
In particular, this process can be constructed as the following: construct a branching process $\{ x_d\}_{d=-m+1}^{m-1}$ such that $x_d=\M_dx$, and, using Theorem \ref{ThDense}, find $\{\w x_d\}_{d=-m+1}^{m-1}\in \cup_{\d>0}\U(\d,m)$ such that
(\ref{xdd}) holds with $\e=\ww\e/(2m-1)$. Then (\ref{xd1}) holds for $\ww x$ selected as the
representative branch of $\{\w x_d\}_{d=-m+1}^{m-1}$.
\end{theorem}
\par
According to Theorem \ref{ThD2}, the set $\cup_{\d>0}U(\d,m)$ is everywhere dense in $\ell_2$; this leads to possibility of applications for sequences from $\ell_2$ of a general kind.

Furthermore, Theorem \ref{ThDeg}  represents an uniqueness result in the sprit of \cite{La,La2,OU08,OU,U1,U2}.
However, the problem of recovery $\ww x$ from its subsequence features some robustness as shown in the following theorem.
\par
Let $B_\d=\{\xi\in\ell_2:\ \|\xi\|_{\ell_2}\le\d\}$, $\d\ge 0$.
\begin{theorem}\label{ThA} Under the assumptions and notations of Theorem \ref{ThDeg},
consider  a problem of recovery of the set $\{\ww x(k)\}_{k=-M}^M$
from a noise contaminated truncated series  of observations  $\{\ww x(mk)+\xi(mk)\}_{k=-N}^N$, where $M>0$ and $N>0$ are integers, $\xi\in\ell_2$ represents a noise contaminating the observations.
 This problem
 is well-posed in the following sense: for any  $M$ and any $\ww\e>0$,
 there exists a recovery algorithm and $\d_0>0$ such that for any $\d\in[0,\d_0)$ and $\xi\in B_\d$ there exists an integer $N_0>0$ such the recovery error $\max_{n=-M,..,M}|\ww x(n)-\ww x(n)|$ is less than $\ww\e$ for all $N\ge N_0$.
Here $\ww x(n)$ is the estimate of $\ww x(n)$ obtained by the corresponding recovery algorithm.
\end{theorem}
\section{Applications to sparse  sampling in continuous time}

Up to the end of this paper, we assume that we are given  $\O>0$ and  $\tau>0$  such that $\O\le\pi/\tau$.
 We will denote $t_k=\tau k$, $k\in\ZZ$.

The classical
Nyquist-Shannon-Kotelnikov Theorem states that {
a band-limited function
$f\in \BLO$ is uniquely defined by the  sequence $\{f(t_k)\}_{k\in \ZZ}$.}

The sampling rate $\tau=\pi/\O$ is called the critical Nyquist rate for $f\in \BLO$.
If $\tau<\pi/\O$, then, for any finite set $S$ or for $S=\ZZ^\pm$,
$f\in\BLO$ is uniquely defined by the values   $\{f(t_k)\}_{k\in\ZZ\backslash S}$, where  $t_k=\tau k$, $k\in\ZZ$; this
was established in \citet{F91,V87}.   We cannot claim the same for some infinite sets of missing values.
For example, it may happen that
$f\in \BLO$  is not uniquely defined by the values $\{f(t_{mk})\}_{k\in\ZZ}$ for $m\in \ZZ^+$, if the sampling rate for this  sample
 $\{f(t_{mk})\}_{k\in\ZZ}$ is lower than is the so-called critical Nyquist rate implied by the Nyquist-Shannon-Kotelnikov Theorem; see more examples in \citet{F95}.
 We address this problem below.

\begin{theorem}\label{ThCTD}  For any $\ww x\in U(\d,m)$, where $\d>0$, there exists
an unique $\ww f\in \BLOO$ such that
$\ww f(t_k)=\ww x(k)$. This
$\ww f$  is uniquely defined by the sequence  $\{\ww f(t_{mk})\}_{k\in \ZZ}$.
\end{theorem}
\par
We say that $\ww f\in F(\d,m)$ is such as described in Theorem \ref{ThCTD}. Theorem \ref{ThCTD} implies  that $F(\d,m)$ can be considered as a class of functions  featuring  spectrum degeneracy of a new kind.

\begin{corollary}\label{corr1} Let $\e>0$ and $f\in \BLOO$ be given.  Let $x\in\ell_2$ be defined as $x(k)=f(t_k)$ for $k\in \ZZ$, and let  $\ww x\in\cup_{\d>0}U(\d,m)$ and $\ww f\in \BLOO$ be defined as described in Theorem \ref{ThCTD}  such that (\ref{xd1}) holds.
 Then this
$\ww f$  is uniquely defined by the values   $\{\ww f(t_{mk})\}_{k\in \ZZ}$, and
\baaa
\|f-\ww f\|_{L_p(\R)}\le C\e,\quad p=2,+\infty.
\eaaa
where $C>0$ depends on $(\O,m,\tau)$ only.
\end{corollary}

\begin{corollary}\label{ThAC} Under the assumptions and notations of Corollary \ref{corr1},
consider  a problem of recovery of the set $\{\ww f(t_n)\}_{n=-M}^M$
from a noise contaminated truncated series  of observations  $\{\ww f(t_{mk})+\xi(mk)\}_{k=-N}^N$, where $M>0$ and $N>0$ are integers, $\xi\in\ell_2$ represents a noise contaminating the observations.
 This problem
 is well-posed in the following sense: for any $M$ and any  $\ww\e>0$,
 there exists a recovery algorithm and $\d_0>0$ such that for any $\d\in[0,\d_0)$ and $\xi\in B_\d$ there exists an integer $N_0>0$ such the recovery error $\max_{n=-M,..,M}|\ww f(t_n)-\ww f_E(t_n)|$ is less than $\ww\e$ for all $N\ge N_0$.
Here $\ww f_E(t_n)$ is the estimate of $\ww f(t_n)$ obtained by the corresponding recovery algorithm.
\end{corollary}
\begin{remark}{\rm
Theorem \ref{ThCTD} and Corollary \ref{corr1} represent uniqueness results in the sprit of \cite{La,La2,OU08,OU,U1,U2}.
 These works considered more general function  with unbounded spectrum domain;  the uniqueness sets of times in these works are not equidistant.
Theorem \ref{ThCTD} deals with sequences rather than with continuous time  time functions  and therefore its result is quite different from  any of results \cite{La,La2,OU08,OU,U1,U2}.
Corollary \ref{corr1} considers band-limited functions only; however, it considers  equidistant uniqueness sets $\{t_{mk}\}_{k\le 0,\ k\in \ZZ}=\{m\tau k\}_{k\le 0,k\in\ZZ}$.
In addition, our proofs are  based on a recovery algorithm featuring some robustness; the papers  \cite{La,La2,OU08,OU,U1,U2} do not consider recovery algorithms and deal  with uniqueness problem.
}\end{remark}
\begin{remark}{\rm
In Corollary  \ref{corr1}, $\ww f$ can be viewed as a result of an arbitrarily small adjustment of $f$.
This $\ww f$ is uniquely defined by sample  with a
sampling distance  $m \tau$, where   $\tau$ is a distance smaller than a the distance defined by  critical Nyquist rate for $f$.
Since the value $\e$ can be arbitrarily small, and $m$ and $N$ can be arbitrarily large in Theorem \ref{corr1} and Corollary \ref{ThA}, one can say  that the restriction on the sampling rate defined by the  Nyquist rate is bypassed, in a certain sense.
}\end{remark}

\subsubsection*{The case of  non-bandlimited continuous time functions}
Technically speaking, the classical sampling theorem is applicable to band-limited continuous time  functions only.
However, its  important role in signal processing is based on applicability  to more general functions since they allow  band-limited approximations: any $f\in L_1(\R)\cap L_2(\R)$ can be approximated by bandlimited functions
$f_\O\defi \F^{-1}(F\Ind_{[-\O,\O]})$ with $\O\to +\infty$, where $F=\F f$, and where $\Ind$ is the indicator function. However, the sampling frequency has to be increased along with $\O$:  for a given $\tau>0$, the sample  $\{f_\O(\tau k)\}_{k\in\ZZ}$
defines the function $f_\O$ if  $\tau\le \pi/\O$. Therefore, there is a problem of representation of general functions via sparse samples. A related problem is aliasing of continuous time processes after time discretization.
Corollary \ref{corr1}  allows to overcome this obstacle in a certain sense as the following.

\begin{corollary}\label{corr2}   For any $f\in L_1(\R)\cap L_2(\R)$,  $\e>0$, and $\Delta>0$, there  exists $\oo\O>0$   such that the following holds for any   $\O\ge \oo\O$:
 \begin{enumerate}
 \item
 $\sup_{t\in\R} |f(t)-f_{\O}(t)|\le\e$, where  $f_{\O}= \F^{-1}F_{\O}$, $F_{\O}=F\Ind_{ [-\O,\O]}$, $F=\F f$.
\item
The function $f_{\O}$ belongs to $\BLO$, and, for $\tau=\pi/\O$, $t_k=\tau k$,   satisfies assumptions of Corollary \ref{corr1}. For this function, for any $\e>0$ there  exists $\ww f\in \BLO$ such that
 $\sup_{t\in\R} |f_{\O}(t)-\ww f(t)|\le\e$ and that $\ww f$  is uniquely defined by the values   $\{\ww f(\t_{k})\}_{k\in \ZZ}$ for an equidistant sequence of sampling points $\t_k =k \Delta$, $k\in\ZZ$, for any $s\in\ZZ$.
\end{enumerate}
\end{corollary}

\section{Proofs}
To proceed with the proof of the theorems, we will need to prove some preliminary lemma first.
\begin{lemma}\label{ThOdd} Let  $x\in \V(\d,\nu m)$ for some $\d>0$ and for $m,\nu\in\ZZ$ such that  $m,\nu\ge 1$. Then the following holds for $\varkappa=1$ or $\varkappa=m$:
\begin{enumerate}
\item
 For any  $n\ge 0$, $n\in\ZZ$, the sequence $\{x(\varkappa k+n)\}_{k\in\ZZ}$
is uniquely defined by the values $\{x{(\varkappa k+n})\}_{k\in Z^-}$.
\item  For any  $n\le 0$, $n\in\ZZ$, the sequence $\{x(\varkappa k-n)\}_{k\in\ZZ}$
is uniquely defined by the values $\{x{(\varkappa k-n})\}_{k\in\ZZ^+}$.
\end{enumerate}
\end{lemma}
\def\dm{{\nu m}}

{\em Proof of Lemma \ref{ThOdd}}.  It suffices to proof the theorem for the case of the extrapolation from the set $\ZZ^-$ only; the extension on the extrapolation from $\ZZ^+$ is straightforward.

Consider a transfer functions and its inverse Z-transform
   \baa
 \w H(z)\defi z^n V(z^{\dm })^n,\qquad \w h=\Z^{-1}\w H, \label{wH}\eaa
where
 \baaa
 V(z)\defi 1-\exp\left[-\frac{\g}{z+ 1-\g^{-r}}\right],
\label{wK}
\eaaa
and where $r>0$ and $\g>0$ are  parameters. This function $V$ was introduced in \cite{D12a}.
(In the notations from \citet{D12a}, $r=2\mu/(q-1)$, where $\mu>1$, $q>1$ are the parameters).
We assume that $r$ is fixed and consider variable $\g\to +\infty$.

In the proof below, we will show that $\w H\ew$ approximates $e^{in\o}$ and therefore defines
a linear $n$-step predictor with the kernel $\w h$.

\par
\def\OO{W}  Let $\a=\a(\g)\defi 1-\g^{-r}$. Clearly, $\a=\a(\g)\to 1$  as $\g\to+\infty$.

Let $\OO(\a)=\arccos(-\a)$, let $D_+(\a)=(-\OO(\a),\OO(\a))$, and let
 $D(\a)\defi[-\pi,\pi]\backslash D_+(\a)$. We have that  $\cos(\OO(\a))+\a=0$, $\cos(\o)+\a>0$ for  $\o\in D_+(\a)$,
and $\cos(\o)+\a<0$ for  $\o\in D(\a)$.
\par

It was shown in \citet{D12a} that the following holds:
\begin{itemize}
\item[(i)] $V(z)\in \HHH^{\infty}(D^c)$ and $zV(z)\in
\HHH^{\infty}(D^c)$.
\item[(ii)] $V(e^{i\o})\to 1$ for all  $\o\in (-\pi,\pi)$ as  $\g\to +\infty$.
\item[(iii)] If $\o\in D_+(\a)$  then $\Re \left(\frac{\g}{e^{i\o}+\a}\right) >0$ and $|V\ew -1|\le 1$.
\end{itemize}

The definitions imply that there exists $\g_0>0$ such that
\baa
\sup_{\o\in(-\pi,\pi]\setminus J_\d}|V\ew-1|\le 1, \quad \forall \g:\ \g>\g_0.
\label{V1}\eaa
Without a loss of generality, we assume below that $\g>\g_0$.

Let
\baaa
&&Q(\a)=\cup_{k=0}^{\dm-1} \left(\frac{\OO(\a)+2\pi k}{\dm},\frac{2\pi-\OO(\a)+2\pi k}{\dm} \right),\qquad \breakk
Q_+(\a)=[-\pi,\pi]\setminus Q(\a).\eaaa
\index{For example, if $m=2$ then
\baaa
&&Q(\a)=(\OO(\a)/2,\pi-\OO(\a)/2)\cup (-\OO(\a)/2,\OO(\a)/2-\pi),\qquad \breakk
Q_+(\a)=[-\pi,\pi]\setminus Q(\a).\eaaa}

\begin{figure}[ht]
\centerline{\epsfig{figure=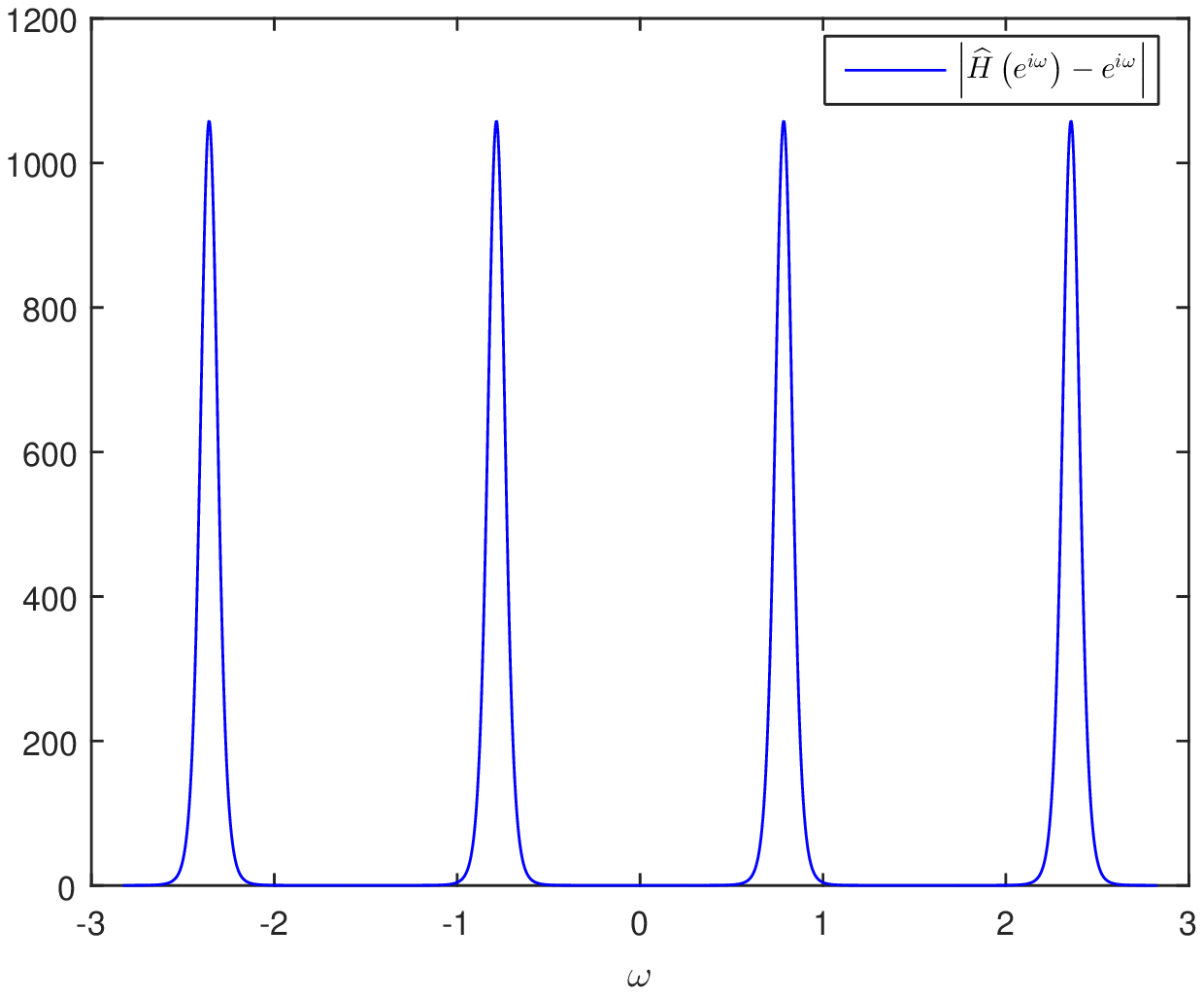,width=9cm,height=6.5cm}}
\centerline{\epsfig{figure=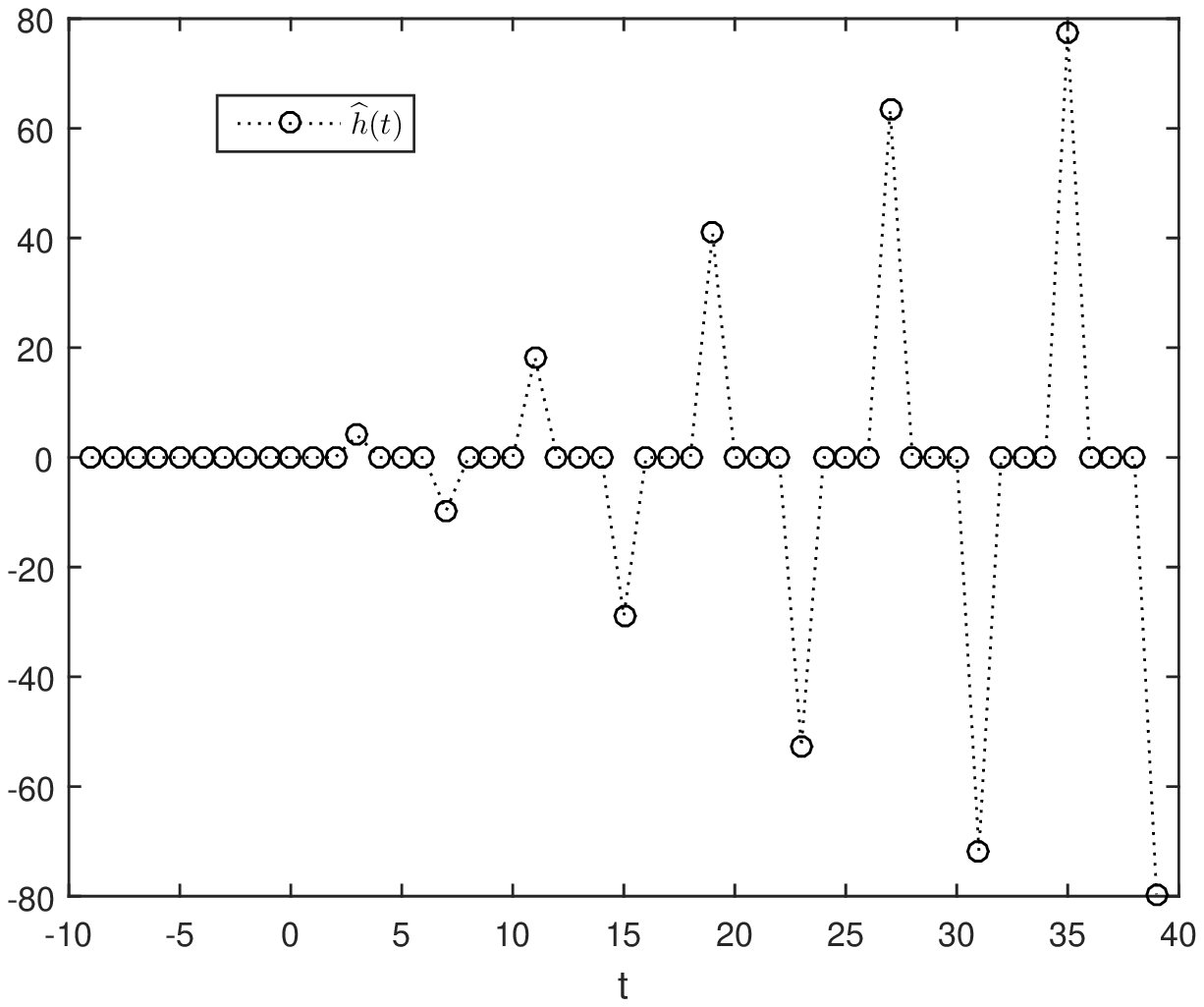,width=9cm,height=6.5cm}}
\vspace{0.5cm} \caption[]{\sm  Approximation of the one-step forward  shift operator: the values
 $|\w H\ew -e^{i\o }|$  for the transfer function of the predictor  (\ref{wH}) and the values of the corresponding kernel $\w h=\Z^{-1}\w H$ with  $\g=4$, $r=0.4$, $m=4$. }
\vspace{0cm}
\label{fig2}
\end{figure}

From the properties of $V$, it follows that the following holds.
\begin{itemize}
\item[(i)] $V(z^{\dm})\in \HHH^{\infty}(D^c)$ and $ zV(z^{\dm})\in
\HHH^{\infty}(D^c)$.
\item[(ii)] $V\left(e^{i \o \dm }\right)\to 1$  and $\w H\ew\to e^{in\o}$ for all  $\o\in (-\pi,\pi]\setminus \{ s_{k,\dm}\}_{k=0}^{\dm-1}$ as  $\g\to +\infty$.
\item[(iii)]
\baaa
\Re \left(\frac{\g}{e^{i \o \dm }+\a}\right) >0,\quad
\left|V\left(e^{i\o \dm }\right) -1\right|\le 1, \quad\brea \o\in Q_+(\a).
\eaaa
\end{itemize}
Figure \ref{fig2} shows an example of the shape of error curves for approximation of
 the  forward one step shift operator. More precisely, they show
 the shape of $|\w H\ew -e^{i\o n}|$  for the transfer function   (\ref{wH}) with $n=\nu=1$,
  and  the shape of the corresponding predicting kernel $\w h$ with some selected  parameters.

By the choice of $V$, it follows that $|V(z)|\to 0$ as $|z|\to +\infty$. Hence  $v(0)=0$ for $v=\Z^{-1}V$ and \baaa
&&V(z^\dm)\breakk =z^{-\dm}v(1)+z^{-2\dm}v(2)+z^{-3\dm}v(3)+.. .\eaaa
 Clearly, we have that
\baaa
V(z^\dm)^n=z^{-n\dm}w(1)+z^{-(n+1)\dm}w(2)+z^{-(n+2)\dm}w(3)+...,\eaaa
where $w=\Z^{-1}(W)$ for $W(z)=V(z)^n$.
Hence \baaa
&&\w H(z^\dm)=z^nV(z^\dm)^n\breakk=z^{n-n\dm}w(1)+z^{n-(n+1)\dm}w(2)+z^{n-(n+2)\dm}w(3)+...\\
&&= z^{n-n\dm}\w h(n\dm-n)+z^{n-(n+1)\dm}\w h((n+1)\dm-n)\breakk+z^{n-(n+2)\dm}\w h((n+2)\dm-n)+...
\eaaa
In particular, $\w H\,\! \in \HHH^{\infty}(D^c)$ and \baa
 \w h(k)=0\quad  \hbox{if either}\quad (k+n)/m\notin \ZZ\quad \hbox{or}\quad  k<mn-n.
 \label{hm}\eaa
(In fact, the sequence $\w h$ is more sparse for $n>1$ or $\nu >1$ than is shown in (\ref{hm}); nevertheless, (\ref{hm}) is sufficient
for our purposes).

It can be noted that $\w h$ is real valued, since  $\w H\,\!\left(\oo z\right)=\overline{\w H\,\!\left(z\right)}$.

\def\ko{h}

Assume  that we are given  $x\in\V(\d,\dm)$. Let $X\defi \Z x$, $Y(z)=zX(z)$, and $\w Y\defi \Z\w y=\w H\,\! X$.
\par
By the notations accepted above, it follows that $B_1$ is the unit ball in $\ell_2$. Let us  show that
\baa
\sup_{k\in\ZZ}|x(k+n)-\w y(k)|\to 0\quad \hbox{as}\quad \g\to+\infty\quad\brea
\hbox{uniformly over}\quad x\in\V(\d,\dm)\cap B_1,
\label{conv}\eaa
 where the process $\w y$ is the output of a linear predictor defined by the kernels $\w h$ as
 \baa
  \w y(k)\defi \sum^{k}_{p=-\infty}\w h(k-p)x(p).\label{predict}
\eaa
This predictor  produces the process $\w y(k)$
approximating $x(k+n)$ as $\g\to +\infty$ for all inputs $x\in \V(\d,\dm)$.

\begin{remark}\label{rem-hkm}{\rm
By (\ref{hm}),(\ref{predict}),  an estimate  $\w y(k)$  of $x(k+n)$ is constructed using  the observations
$\{x(k+n-qm)\}_{q\ge n, q\in\ZZ}$. This can be seen from representation of convolution (\ref{predict}) as
\baaa
y(k)=\sum_{j\in\ZZ,j\ge n}\w h(jm-n) x(k-jm+n).
\eaaa
}
\end{remark}

 \par Let $\yo (t)\defi x(t+n)$ and $Y=\Z y$. We have that $\| Y\ew-\w Y\ew\|_{L_1(-\pi,\pi)}=I_1+I_2,$
where \baaa &&I_1=\int_{Q(\a)}| Y\ew-\w Y\ew| d\o,\qquad\breakk
I_2=\int_{Q_+(\a)}|Y\ew-\w Y\ew| d\o. \eaaa

By the definitions, there exists  $\g_0>0$ such that, for all  $\g>\g_0$, we have that \baa
 Q(\a)\subset J_{\d,m},\quad I_1=0.\label{alga}\eaa
 This means that $I_1\to 0$ as $\g\to +\infty$
 uniformly over $x\in\V(\d,\dm)\cap B_1$.

Let us estimate $I_2$. We have that
\baaa I_2=\int_{Q_+(\a)} | (e^{i\o n}-H\,\!\left(e^{i\o  }\right)) X\ew|  d\o\break\le
\int_{Q_+(\a)} |e^{i\o n}(1-V\left(e^{i\o \dm }\right)^n) X\ew|  d\o\\\le
\|1-V\left(e^{i\o \dm }\right)^n\|_{L_2(Q_+(\a))} \|X\ew\|_{L_2(-\pi,\pi)}.\eaaa
\par
Further,
$\Ind_{Q_+(\a)}(\o) |1-V\left(e^{i\o\dm}\right)^n| \to 0$ a.e. as $\g\to
+\infty$. By (\ref{V1}), \baaa
\Ind_{Q_+(\a)}(\o)|1-V\left(e^{i\o \dm }\right)^n | \le 2^n, \quad \forall \g:\ \g>\g_0.\eaaa   From Lebesgue Dominance Theorem,
it follows that
$\|1-V\left(e^{i\o \dm }\right)^n\|_{L_2(Q_+(\a))}\to 0$ as $\g\to+\infty$. It follows that
$I_1+I_2\to 0$ uniformly over $x\in\V(\d,\dm)\cap B_1$.  Hence (\ref{conv}) holds and
\baaa &&\sup_{k\in\ZZ}|x(k+n)-\w y(k)|\to 0\quad \hbox{as}\quad \g\to+\infty\quad\breakk
\hbox{uniformly over}\quad x\in\V(\d,\dm)\cap B_1.
\label{conv2}\eaaa
   Hence the
predicting kernels $\w h=\Z^{-1}\w H\,\!$ are such as required.

To complete the proof of    Lemma \ref{ThOdd}, it suffices to
prove that if $x$ is such that $x(k\varkappa +n)=0$ for
$k\le s$, then $x(k\varkappa +n)=0$ for $k>q$ for all $q\in\ZZ$.
By the predictability established by (\ref{conv})  for a process $x\in\V(\d,\dm)$ such that
$x(k\varkappa +n)=0$ for $k\le s$, it follows that $x(k\varkappa+\varkappa+n)=0$;  we obtain this by letting $\g\to +\infty$
in (\ref{wH}). (See also Remark \ref{rem-hkm}). Similarly, we obtain that $x(k\varkappa+2\varkappa+q)=0$.
Repeating this procedure $n$,
we obtain that $x(k\varkappa+n)=0$ for all $k\in\ZZ$.
This
completes the proof of Lemma \ref{ThOdd}. $\Box$

\begin{remark} \label{lemmaAD} {\rm Assume that $\{\w x_d\}_{d=-m+1}^{m-1}\in \U(\d,m)$ for some $\d>0$.
\begin{enumerate}
\item Lemma \ref{ThOdd} applied with $\varkappa=1$ implies  that all $\w x_d$  are uniquely defined by $\{\w x_0(k)\}_{k\in\ZZ^-}$  for  $d=0,1,...,m-1$.
\item Lemma \ref{ThOdd} applied with $\varkappa=m$ and $n=0$ implies  that all values $\w x_d(mk)$ for $k>0$ are uniquely defined by  the sequence $\{\w x_0(mk)\}_{k\in\ZZ^+}$  for  $d=-m+1,1,...,1$.
\end{enumerate}
}\end{remark}

{\em Proof of Theorem \ref{ThDeg}}.
Let us prove statements (i)-(ii) for $s=0$.
 Let $d\in\{0,1,...,m-1\}$.
By Lemma \ref{ThOdd} applied with $\varkappa=m$, we have that
 a sequence  $\{\w x_d(mk)\}_{k\ge 1}$
is uniquely defined by
the sequence  $\{\w x_d(mk)\}_{k\le 0}$ (see Remark \ref{rem-hkm} above).
Since $\{\ww x(mk-d)\}_{k\ge 1}=\{\w x_d(mk)\}_{k\ge 1}$ and  $\{\w x_d(mk)\}_{k\le 0}=\{\ww x(mk)\}_{k\le 0}$, this implies that both sequences $\{\ww x(mk-d)\}_{k\ge 1}$ and $\{\w x_d(mk)\}_{k\ge 1}$
are uniquely defined by
the sequence  $\{\ww x(mk)\}_{k\le 0}$.
This completes statements (i) for the case where $s=0$.
\par
Similar reasoning can be applied for statement (ii). Let $d\in\{-m+1,...,0\}$.
By Lemma \ref{ThOdd} applied with $\varkappa=m$ again, we have that
 a sequence  $\{\w x_d(mk)\}_{k\le -1}$
is uniquely defined by
the sequence  $\{\w x_d(mk)\}_{k\ge 0}$.
Since $\{\ww x(mk-d)\}_{k\le -1}=\{\w x_d(mk)\}_{k\le -1}$ and  $\{\w x_d(mk)\}_{k\ge 0}=\{\ww x(mk)\}_{k\ge 0}$, this implies that both sequences $\{\ww x(mk-d)\}_{k\le -1}$ and $\{\w x_d(mk)\}_{k\le -1}$
are uniquely defined by
the sequence  $\{\ww x(mk)\}_{k\ge 0}$.
This completes statements (ii) for the case where $s=0$.
Further, statements (i)-(ii) for $s\neq 0$ follows from the corresponding
statements for $s= 0$ applied for the sequences  with shifted time $\{\ww x(k+s)\}$ and $\{\w x_d(k+s)\}$.   Statements (i)-(ii) imply statement (iii).
This completes the proof of Theorem \ref{ThDeg}. $\Box$
\par
{\em Proof for Example \ref{ex1}}.  It suffices to show that
  $s_{\mu(m,d_1),k}\neq s_{\mu(m,d_2),l}$ if $d_1\neq d_2$ for all $k,l\in\ZZ$.
 Suppose that
$s_{\mu(m,d_1),k}= s_{\mu(m,d_2),l}$ for some $k,l\in\ZZ$ for $d_2=d_1 +r$, for some
 $d_1,d_2\in\{-m+1,...,m-1\}$ such that $r\defi d_2-d_1> 0$. In this case, the definitions imply that
\baaa
 \frac{2 k-1}{ 2^{d_1}}=\frac{2 l-1}{2^{d_2} }.
 \eaaa
This means that  $ 2^r(2 k-1)=2 l-1$,  and, therefore the number $2^r(2 k-1)$ is odd. This is impossible since we had assumed that  $r>0$. Hence
 the sets $J_{\d,\mu(m,d)}$ are disjoint for different
 $d$ for small $\d>0$.  This completes the proof for Example \ref{ex1}. $\Box$

\par
{\em Proof of Theorem \ref{ThDense}}.  Without a loss of generality, we assume that all  $\d$ used below are small enough to ensure that the sets $J_{\d,\mu(m,d)}$ are disjoint for different
 $d$.
\par
Let $Y_d\defi X_0-X_d$ and $y_d=\Z^{-1}Y_d$, where $ X_d=\Z  x_d$, $d=-m+1,...,m-1$.
Since $x_d(k)=x_0(k)$ for $k\le 0$ and $d>0$, and $x_d(k)=x_0(k)$ for $k\ge  0$ and $d<0$, it follows that $y_d(k)=0$ for $k\le 0$ and $d>0$, and $y_d(k)=0$ for $k\ge 0$ and $d<0$.
 $d=1,2,...,m-1$. In addition, $Y_0\equiv 0$ and $y_0\equiv 0$.

Further, let $\TT=\{k\in\ZZ:\ |k|\le m-1,\ k\neq 0\}$, let
\baaa
\w X_0\ew\defi  X_0\ew\Ind_{\{\o\notin \cup_{d=-m+1}^{m-1} J_{\d,\mu(m,d)}\}}\brea +
\sum_{d\in\TT} Y_d\ew\Ind_{\{\o\in J_{\d,\mu(m,d)}\}},\eaaa
and let $\w X_d\defi \w X_{0}-Y_d$ for $d\in\TT$. By the definitions,
\baaa
\w X_d\ew=\w X_{0}\ew-X_0\ew+X_d\ew\brea=X_d\ew+  \sum_{p\in\TT} (Y_p\ew-X_0\ew)\Ind_{\{\o\in J_{\d,\mu(m,p)}\}}\\=X_d\ew- \sum_{p\in\TT}X_p\ew\Ind_{\{\o\in J_{\d,\mu(m,p)}\}}\\=X_d\ew\Ind_{\{\o\notin J_{\d,\mu(m,d)}\}}- \sum_{p\in\TT, p\neq d}X_p\ew\Ind_{\{\o\in J_{\d,\mu(m,p)}\}},
 \quad\\ d=-m+1,...,m-1.
 \eaaa
Since the sets $J_{\d,\mu(m,d)}$ are mutually disjoint,
 it follows that   $\w X_d\ew =0$ for
$\o\in J_{\d,\mu(m,d)}$ and for all $d$. Hence $\w x_d\defi\Z^{-1}\w X_d\in\V(\d,\mu(m,d))$
for
all $d$. It follows that the branching  process
$\{\w x_d\}_{d=-m+1}^{m-1}$ belongs to $\cup_{\d>0} \U(\d,m)$.
In particular, $\w x_d(k)=\w x_0(k)$, for $k\le 0$ and
 $d>0$, and  $\w x_d(k)=\w x_0(k)$, for $k\ge 0$ and
 $d<0$. Clearly, $\|\w x_d-x_d\|_{\ell_2}\to 0$ as $\d\to 0$ for $d=-m+1,,...,m-1$.
This completes the proof of Theorem \ref{ThDense}. $\Box$

{\em Proof of Theorem \ref{ThD2}}.
Let  $\{ x_d\}_{d=-m+1}^{m-1}$, $\{\w x_d\}_{d=-m+1}^{m-1}\in \cup_{\d>0}\U(\d,m)$ be such that
suggested in the theorem's  statement.  By the definitions, it follows that
\baaa
\ww x(k)-x(k)=\w x_d(k+d)-x_d(k+d),\quad \brea k\ge 0,\quad d\ge 0,\quad (k+d)/m\in\ZZ,
\eaaa
and
\baaa
\ww x(k)-x(k)=\w x_d(k+d)-x_d(k+d),\quad \brea k<0,\quad d<0,\quad (k+d)/m\in\ZZ.
\eaaa
Then (\ref{xd1}) follows from
(\ref{xdd}). This completes the proof of Theorem \ref{ThD2}. $\Box$
\par
{\em Proof of Theorem \ref{ThA}.}
It suffices to show that the error for recovery a singe term $\ww x(n)$ for a given integer $n$  from the sequence $\{\ww x(mk)\}_{k\in \ZZ,\, k\le 0}$ can be made arbitrarily small is a well-posed problem; the proof for a finite set of values  to recover is similar.
Furthermore, it suffices to consider $n>0$; the case of $n<0$ can be considered similarly.

Let us consider an input sequence $x\in\ell_2$
such that \baa
x=\ww x+\eta,\label{xeta}
\eaa
where $\eta\in\ell_2$ represents a
noise, and where $\ww x\in U(\d,m)$ is the representative branch for a branching process
$\{\w x_d\}_{d=-m+1}^{m-1}\in\U(\d,m)$.

Let $\eta_n\in\ell_2$ be defined such that $\eta_n(k)=\eta(k) \Ind_{\{k\le n\}}$.
Let $N\defi \Z\eta_n$, and let $\s=\|N\ew\|_{L_2(-\pi,\pi)}$; this parameter  represents the
intensity of the noise. Let  $X=\Z x$, $\w X_d=\Z \w x_d$, and $N=\Z \eta$.

Let $d\in\{0,1,...,m-1\}$ be such that $(n+d)/m\in\ZZ$.

Let $\d>0$ be given and an  arbitrarily small $\e>0$ be given.
Assume that  the parameters $(\g,r)$ of $V$ and $\w H$ in (\ref{wH})
are selected such that \baa
\sup_{k\in\ZZ}|\oo x(k+n)-\oo y(k)|\le \e/2 \quad \forall  \oo x\in \V(\d,\mu(m,d))\cap B_1,
\label{yy}\eaa
for  $\oo y=\Z^{-1}(\w H\oo X)$ and $\oo X=\Z\oo x$. Here $\w H$ is selected by (\ref{wH}), and $B_1$ is the unit ball in $\ell_2$.

  By the choice of $d$, we have that there exists $p\in\ZZ$, $p\ge 1$, such that $n=pm-d$ and
\baa
\w x_d(pm)=\ww x(pm-d)=\ww x(n).
\label{d}
\eaa
By (\ref{hm}) and Remark \ref{rem-hkm}, the kernel $\w h$ produces an estimate $y_d$  of
$\w x_d(pm)$ based on observations of $\{\w x_d(km)\}_{k< 0}$.

Let us assume first that $\s=0$. In this case,  we have that
\baa \w E\defi  |\w y_d(0)-\w x_d(pm)|= |\w y_d(0)-\ww x(n)|\le\e/2.\label{eps}\eaa

Let us consider  the case where $\s>0$.  In this case, we have
 that \baaa
  |\w y_d(0)-\w x_d(pm)|= |\w y_d(0)- x(n)|\le \w E+ E_{\eta},\eaaa
where
\baaa
E_{\eta}\defi \frac{1}{2\pi}\|(\w H\ew-e^{i\o n})N\ew|\|_{L_1(-\pi,\pi)}\le \s (\kappa+1),
\eaaa
and where
\baaa
\kappa\defi \sup_{\o\in[-\pi,\pi]}|\w H\ew|.\eaaa
\par
Assume that   $\eta(k)= \xi(k)\Ind_{\{-N\le k\le n\}}-\w x_0(k)\Ind_{\{k<-N\}}$, for an integer $N>0$.
In this case, (\ref{xeta}) gives that
$x(k)=(x_0(k)+\xi(k))\Ind_{\{k>-N\}}$ for $k\le n$. In addition, we have in this case that  $\s\to 0$ as $N\to +\infty$ and $\|\xi\|_{\ell_2(-N,0)}\to 0$. If $N$ is large enough and  $\|\xi\|_{\ell_2(-N,0)}$ is small enough  such that $\s (\kappa+1)<\e/2$, then
$\|\w y_d(0)-   x(n) \|_{\ell_{\infty}}\le\e$.
This completes the proof of Theorem \ref{ThA}. $\Box$

\begin{remark}{\rm
By the properties of $\w H$, we have that $\kappa\to +\infty$  as $\g\to +\infty$.
 This implies that error (\ref{eps}) will be increasing if $\w\e\to 0$ for any given $\s >0$.
This means that, in practice, the predictor  should not  target too small a size of the
error, since in it impossible to ensure that $\s=0$ due inevitable data truncation.
 }\end{remark}

{\em Proof of Theorem \ref{ThCTD}}.  The previous proof shows that $\ww x\in\ell_2$.
For $\ww X=\Z \ww x$,  we have that
\baaa
\ww x(k)=\frac{1}{2\pi}\int_{-\pi}^{\pi}
\ww X\left(e^{i\nu}\right) e^{i\nu k}d\nu\brea =\frac{\tau}{2\pi}\int_{-\pi/\tau}^{\pi/\tau}\ww X\left(e^{i\tau\o}\right) e^{i\o\tau k}d\o,
\eaaa
with the change of variables $\nu=\tau \o$. Let us define function $\ww F: i\R\to \C$ as  $\ww F(i\o)\defi\tau \ww X\left(e^{i \tau\o  }\right)$ for $\o\in\R$.
Then
\baaa
\ww x(k)=\frac{1}{2\pi}\int_{-\pi/\tau}^{\pi/\tau}
\ww F\left(i\o\right) e^{i\o\tau k}d\o.
\eaaa
 Since $\ww X\left(e^{i\cdot}\right)\in L_2(-\pi,\pi)$, this implies   that $\ww F\left(i\cdot\right)\in L_2(i\R)$.
The  sequence $\{\ww x(k)\}_{k\in \ZZ}$ represents the sequence of Fourier coefficients of $\ww F$ and defines  $\ww F$ uniquely. By Theorem \ref{ThDeg}, this sequence is uniquely defined
 by the sequence $\{\ww x(mk)\}_{k\in \ZZ}$.  Let $\ww f\defi \F \ww F$. Clearly,  $\ww f\in \BLOO$ and it is uniquely defined
 by the sequence $\{\ww x(mk)\}_{k\in \ZZ}$.
This completes the proof of Theorem \ref{ThCTD}.
$\Box$

{\em Proof of Corollary \ref{corr1}}.    We use notations from the proofs above.
The existences of the required $\{\w x _d\}_{-m+1}^{m-1}$, $\ww x$, and $\ww f$,  follows from Theorem \ref{ThDeg}-\ref{ThDense}.
We have for $F=\F  f$ that, for some $\oo C>0$,  that
\baaa
&&\|\ww F(i\cdot)-F(i\cdot)\|_{L_2(\R)}= \sqrt{\tau}\|\ww x-x\|_{\ell_2} \breakk \le
\sqrt{\tau}\sum_{d=-m+1}^{m-1}\|\w x_d-x_d\|_{\ell_2}\le  \sqrt{\tau}(2m-1) \e.
\eaaa
Since $F(i\o)=0$ and $\ww F(i\o)=0$ if $|\o|>\pi/\tau$,
it follows that \baaa
\|\ww F(i\cdot)-F(i\cdot)\|_{L_1(\R)}\le \sqrt{2\O}\|\ww F(i\cdot)-F(i\cdot)\|_{L_2(\R)}.
\eaaa
Combining these estimates, we obtain  that
 \baaa
 \|\ww f-f\|_{L_p(\R)}\le  2\pi\max(1,\sqrt{2\O})\|\ww F(i\cdot)-F(i\cdot)\|_{L_2(\R)}\brea \le
2\pi\max(1,\sqrt{2\O})\sqrt{\tau}(2m-1) \e .
 \eaaa
This completes the proof of  Corollary \ref{corr1}. $\Box$
\par
{\em Proof of Corollary \ref{ThAC}.} It suffices to observe that $\ww x(k)=\ww f(t_k)$
satisfy the assumptions of of Corollary \ref{ThAC}. $\Box$

\par
 {\em Proof of Corollary \ref{corr2}.}
 We have that
 \baaa
&&\sup_{t\in\R}|f(t)-f_{\O}(t)|\le \frac{1}{2\pi}\int_{-\infty}^\infty |F(i\o)- F_{\O}(i\o)|d\o\breakk=\frac{1}{2\pi}\int_{\R\setminus [-\O,\O]} |F(i\o)|d\o
 \to 0\quad\\ &&\hbox{as}\quad \O\to +\infty,\ m\to+\infty.\eaaa
 Then statement (i) follows. The function $f_{\O}$ has the same properties as the function $f$ in Corollary \ref{corr1}. This completes  the proof of Corollary \ref{corr2}.  $\Box$
%
\def\NN{\eta}
\xxxonly{ \section{On numerical implementation}
\label{secA}
 Corollary \ref{corrAC} and Theorem \ref{ThA} allow to bypass, in a certain sense, the restriction on the sampling rate defined by the critical   Nyquist rate.  Admittedly, this is only the very first step in solution of the problem.
The proofs of the results implies a  recovering algorithm. However,
 numerical implementation to data compression and recovery using the scheme
in the proof above  is numerically  challenging.  So far, this possibility is rather theoretical, since
 the presence of a noise in the data or in the measurements can cause a significant error
 if $\g\to +\infty$. The aims of the present paper are limited by the theoretical aspects of possibility of recovering
 functions from decimated samples; we leave analysis of  possibilities for numerical implementation for the future research. However, let us summarize briefly some steps require for the numerical implementation.
\subsubsection*{An algorithm for compression and recovery}
Assume that we are given  $f\in L_1(\R)\cap L_2(\R)$. Corollary  \ref{ThAC}
and Corollary \ref{corr2} offer a method of compressed representation of $f$  via  a sparse sample of some function $\ww f$ that is close enough to $f$; this can be considered as a compression problem.

As is described in Corollary \ref{corr2},  $f$ can be approximated by bandlimited functions
$f_\O\defi \F^{-1}(F(i\o)\Ind_{[-\O,\O]}(\o))$, $F=\F f$, where $\O\to +\infty$.
The classical Sampling Theorem allows  to
restore the Fourier transform $f_\O$ from the sampling series $\{f_\O(t_k)\}_{k\in\ZZ}$, where
$t_k=\tau k$, $k\in\ZZ$, $\tau\in(0,\pi/\O)$. The  aim is a compressed representations
using  the sampling times $mt_k$, where $m>0$ defines a desired sparsity. For this, we suggest to approximate $f_\O$ by a function $\ww f$ such as described in Corollary \ref{corr1}.
The corresponding  steps for this compression algorithm can be summarized as follows.
 \begin{itemize}
 \item[(C1)]  Select $\O>0$ to ensure sufficient approximation of $f$ by $f_\O$ described in Corollary \ref{corr2}.
  Select $\tau\le \pi/\O$ and define $x_0(k)=f(t_k)$, where $t_k=k\tau$, $k\in \ZZ$.
  Select a integer $m>0$ such that $m\tau k$ is a  distance that ensures sufficient sparsity.
  \item[(C2)] Define $x_d=\M_dx_0$, where $d=-m+1,...,m-1$.
 \item[(C3)] Define $\w X_d$ as described in the proof of Theorem \ref{ThDense} using sufficiently small $\d>0$.
 Find $\w x_d=\Z^{-1}\w X_d$; these sequences form a branching process $\{\w x_d\}_{d=-m+1}^{m-1}$. Find the representative branch  $\ww x$
  by Definition  \ref{defMC}. \end{itemize}
Any of steps (C1)-(C3) represents a well-posed problem.

The sequence   $\{\ww x(mk)\}_{k\in \ZZ}$ can be accepted as a compressed representation of $f$.
This can be justified as the following.
The sequence   $\ww x$ is uniquely   defined by the sequence $\{\ww x(mk)\}_{k\in\ZZ}$, and there exists an unique $\ww f\in \BLOO$ such  that  $\ww f(t_{k})=\ww x(k)$ for all $k\in\ZZ$.    This  $\ww f$ approximates $f_\O$ given that $\d$ is sufficiently small. In addition,
 $\ww f$ is uniquely defined by $\ww x$ and hence by $\{\ww x(mk)\}_{k\in\ZZ}$.

Recovery $\ww f$ from  the suggested  compressed representation $\{\ww x(mk)\}_{k\in \ZZ}$ would require the following steps.
 \begin{itemize}
 \item[(R1)] Recover $\ww x$ from the sequence $\{\ww x(mk)\}_{k\in\ZZ}$.
\item[(R2)]    Recover $\ww f$ from $\ww x$.
\end{itemize}

Step (R2) represents a  well-posed problem which solution is implied by the Sampling Theorem. In addition, Theorem \ref{ThA}  establishes some well-posedness of Step (R1) for
recovery of a finite set of values.
}

\section{Discussion and future developments} The paper shows that recoverability of sequences from their decimated subsequences
is associated with certain spectrum degeneracy of a new kind, and that a
sequences of a general kind can be approximated by sequences featuring this degeneracy.
   This is applied to sparse sampling of continuous time band-limited functions.
The paper suggests a uniqueness result for  sparse sampling (Theorems \ref{ThD2} and Theorem \ref{corr1}), and establishes some  robustness of recovery
(Theorems \ref{ThA} and Theorem \ref{ThAC}). This means that the restriction on the sampling rate defined by the  Nyquist rate the Sampling Theorem is bypassed, in a certain sense. This was only the very first step in attacking  the problem; the  numerical implementation to data compression and recovery is quite challenging and there are many open questions. In particular, it involves the solution of ill-posed inverse problems. In addition, there are other possible extensions of this work that we will leave for the future research.
\begin{enumerate}
\item
A theoretical problem arises: {\em How to detect a trace $x_0|_{k\le 0}$ that can be extended
into a branching process featuring degeneracy  described in Definition \ref{defAD} for $d>0$?}
This is actually a non-trivial question even for $m=1$; see discussion Definition 1 in \cite{DX16b}
and discussion in \cite{Df,DX16b,D17b}.
\item To cover applications related to  image processing, the approach of this paper has to be extended  on  2D sequences and functions $f:\R^2\to\R$.
This seems to be a quite difficult task; for instance, our approach is based on the predicting algorithm \cite{D12a} for 1D sequences, and it is unclear if it
can be extended on processes defined on multidimensional lattices.  Possibly, the setting from \citet{PM} can be used.
\item The result of this paper allows many acceptable modifications such as the following.
 \subitem -- The choice of mappings $\M_d$ allows many modifications;  for example,
 $x_d=\M_d x$ are defined such that  there exists $\t\ge m$ such that
$x_d(k)=x(k-d)$ for $k>\t$, without any restrictions on $\{x_d(k)\}_{k=1,..,\t}$.
\subitem --    Conditions of Theorem \ref{ThOdd} can be relaxed: instead of the condition that the spectrum  vanishes  on open sets $J_{\d,m}$, we can require that the  spectrum  vanishes only at the middle points of the intervals forming $J_{\d,m}$; however, the rate of vanishing has to be sufficient, similarly to \citet{D12a}. This is because the predicting algorithm \citet{D12a} does not  require that the spectrum of an underlying process is vanishing on an open subset of $\T$.

\subitem --  The choice (\ref{wH}) of predictors presented in the proofs above is not unique. For example,
we could use  a predicting algorithm from \citet{D12b}  instead of the
the algorithm based on \citet{D12a} used above.
Some examples of numerical experiments for the predicting algorithm based on the transfer function (\ref{wH}) \xxxonly{assumed by Step (R1)}  can be found in  \citet{D16} (for the case where $m=n=1$, in the notations of the present papers).
 \end{enumerate}

{\subsection*{Acknowledgment} This work  was
supported by ARC grant of Australia DP120100928 to the author.}



 \end{document}